\documentclass[10pt,aps,prb,twocolumn,noshowpacs]{revtex4-2}
\usepackage{amsmath,graphicx,amsbsy,amssymb,amsmath,epsfig,latexsym,wasysym,pifont,mathrsfs,natbib}
\usepackage{float} \newfloat{widefig}{thp}{lop} \usepackage{comment}

\usepackage{array, makecell} \usepackage{verbatim} \usepackage{datetime}
\usepackage{multirow,array} \usepackage{hyperref} \usepackage{color} \usepackage{setspace}
\usepackage{gensymb}
\usepackage{subcaption}
\usepackage{graphicx}
\usepackage[export]{adjustbox}

\hypersetup{colorlinks,
  linkcolor=blue,%
  citecolor=blue,%
  urlcolor=blue}

\begin{document}
\title{\textbf{Consequences of magneto-electrical coupling in multiferroic VSe$_{2}$$/$Sc$_{2}$CO$_{2}$ heterostructures.}}

\author{Himangshu Sekhar Sarmah}
\email[]{shimangshu@iitg.ac.in}
\affiliation{Department of Physics, Indian Institute of Technology
  Guwahati, Guwahati-781039, Assam, India.}    
\author{Subhradip Ghosh}
\email{subhra@iitg.ac.in} \affiliation{Department of Physics,
  Indian Institute of Technology Guwahati, Guwahati-781039, Assam,
  India.} 
\begin{abstract}
	Two-dimensional van der Waals heterostructures are potential game changers both in understanding the fundamental physics and in the realization of various devices that exploit magnetism at the nanoscale. Multiferroic heterostructures comprising a two-dimensional ferroelectric and a two-dimensional ferromagnet are ideal candidates for electrical control of properties of the ferromagnets that can lead to non-volatile memory devices, for example. Relatively new but immensely promising two-dimensional materials, MXene and transition metal dichalcogenides, can be effectively combined to achieve the goal as both have flexibilities in their structures and compositions that are tunable. In this work, using Density Functional Theory, we have investigated the magneto-electric coupling driven transitions in the electronic ground states of VSe$_{2}$-Sc$_{2}$CO$_{2}$ bi-layer and tri-layer heterostructures. Our results demonstrate that the change in the ferroelectric polarisation in the MXene layer leads to changes in the spin-polarized band structures of the magnetic component VSe$_{2}$ enabling a semiconductor to half-metal transition in these heterostructures. We propose several applications of this magneto-electric coupling in these multiferroic heterostructures that can lead to the efficient operation of Field Effect transistors and achieve non-volatility in memory devices at the nanoscale.
\end{abstract}

\maketitle

\section{Introduction\label{intro}}
Lately, electronic devices for information and communications are in high demand as the storage of huge amounts of data and faster communication are the keys to success in an era of data-driven science and technology. Non-volatility of the hardware components is one basic requirement in this scheme so that a large cache of data can be retained every time the device is disconnected from the power supply \cite{spintronics}. With the advent of storage and various electrical devices that exploit the science of magnetism, remarkable advances have been possible in this direction \cite{sdsarma,oxidenano,ultraspin,spintronics1,spintronics2,spintronics3,spintronics4}. At the same time, demand for miniaturization of the devices has led to the exploration of materials for devices in low dimensions. The discovery of magnetism in two-dimensions (2D) \cite{huang2017layer,gong2017discovery,gronke2019chromium,wang2018electric,sun2020room,gong2019experimental} has given a substantial boost towards the realization of miniaturized magnetic storage and various electrical devices. At the heart of this is the electrical control of atomic-thick Ferromagnetism, in particular. The idea has been explored in cases of prototype 2D magnets like CrI$_{3}$ \cite{cri3,cri31,cri32}, Cr$_{2}$Ge$_{2}$Te$_{6}$ \cite{crgete} and Fe$_{2}$GeTe$_{2}$ \cite{fegete,fegete1}. Though promising, exploring them as non-volatile memory or storage devices is bound to fail as they require continuous electrical control for the sustenance of the states induced by electrical switching.

The non-volatile electrical control of magnetic devices can be realized in multiferroic materials where the magnetism is controlled by the Ferroelectric polarization. Since Ferroelectric materials are endowed with bi-stable polarized states, their electrical switching is completely non-volatile. Thus, if the magnetic properties can be coupled with the states of electric polarization, the resulting magneto-electric coupling will lead to non-volatile electrical control of magnetic states \cite{dong2019magnetoelectricity,gong2019multiferroicity,sun2019valence}. However, single-phase multiferroics are not very attractive for applications as none of the existing materials combine large and robust electric and magnetic polarizations at room temperature \cite{sciencearticle}. This has been circumvented by forming two-phase composites with a Ferroelectric and a Ferromagnetic component. The magneto-electric effect is significant if the coupling at the interface of the two is large. This is more effective if the area at the interface is large.

Making 2D Van der Waals (vdW) heterostructures with a 2D Ferroelectric and a 2D ferromagnet, thus, can be a useful way to achieve this. With the synthesis of several 2D Ferroelectrics \cite{wan2019nonvolatile,wang2020exploring,yu2018unraveling,chang2016discovery,liu2016room,noor2023engineering}, construction of such heterostructures no longer remains a distant reality. Very recently, this idea has been demonstrated to be working as a proof of concept through Density Functional Theory \cite{dft} based calculations \cite{zhao2019nonvolatile,liu2023nonvolatile,wu2024nonvolatile,cao2022multiferroic,vse2,maz}. In those calculations, MXene Sc$_{2}$CO$_{2}$, recently predicted to be possessing large out-of-plane Ferroelectric polarization \cite{sc2co2abhishek} from DFT calculations, is considered to be the Ferroelectric component in the 2D heterostructure. Different 2D Ferromagnets having minimal lattice mismatch with Sc$_{2}$CO$_{2}$ have been used as the magnetic component. The calculations showed that the electronic ground states of the heterostructure can be altered between semiconducting and half-metallic upon changes in the Ferroelectric polarization. Taking a cue from these recent developments, in this work, we have explored the consequences of magneto-electrical coupling in bi-layer and tri-layer VSe$_{2}$/Sc$_{2}$CO$_{2}$ heterostructures. The transition-metal dichalcogenide(TMDC)-MXene heterostructures have been considered as both have flexibilities in their compositions and structures that can be exploited to realize tunable functional properties. As the magnetic component of the heterostructures, we have considered the 2H phase of VSe$_{2}$ as its monolayer has been synthesized recently \cite{you2022salt}. Our results show that the electronic ground states of these heterostructures can be reversibly switched between semiconducting and half-metallic states upon switching electrical polarization of Sc$_{2}$CO$_{2}$. Based on these findings, we propose these heterostructures for multiple applications like bi-polar magnetic semiconductors, tunneling field effect transistors, and low-cost, power-efficient non-volatile memory devices.

\section{Computation Details}
\begin{figure*}
    \includegraphics[height=12cm, width=18.00 cm]{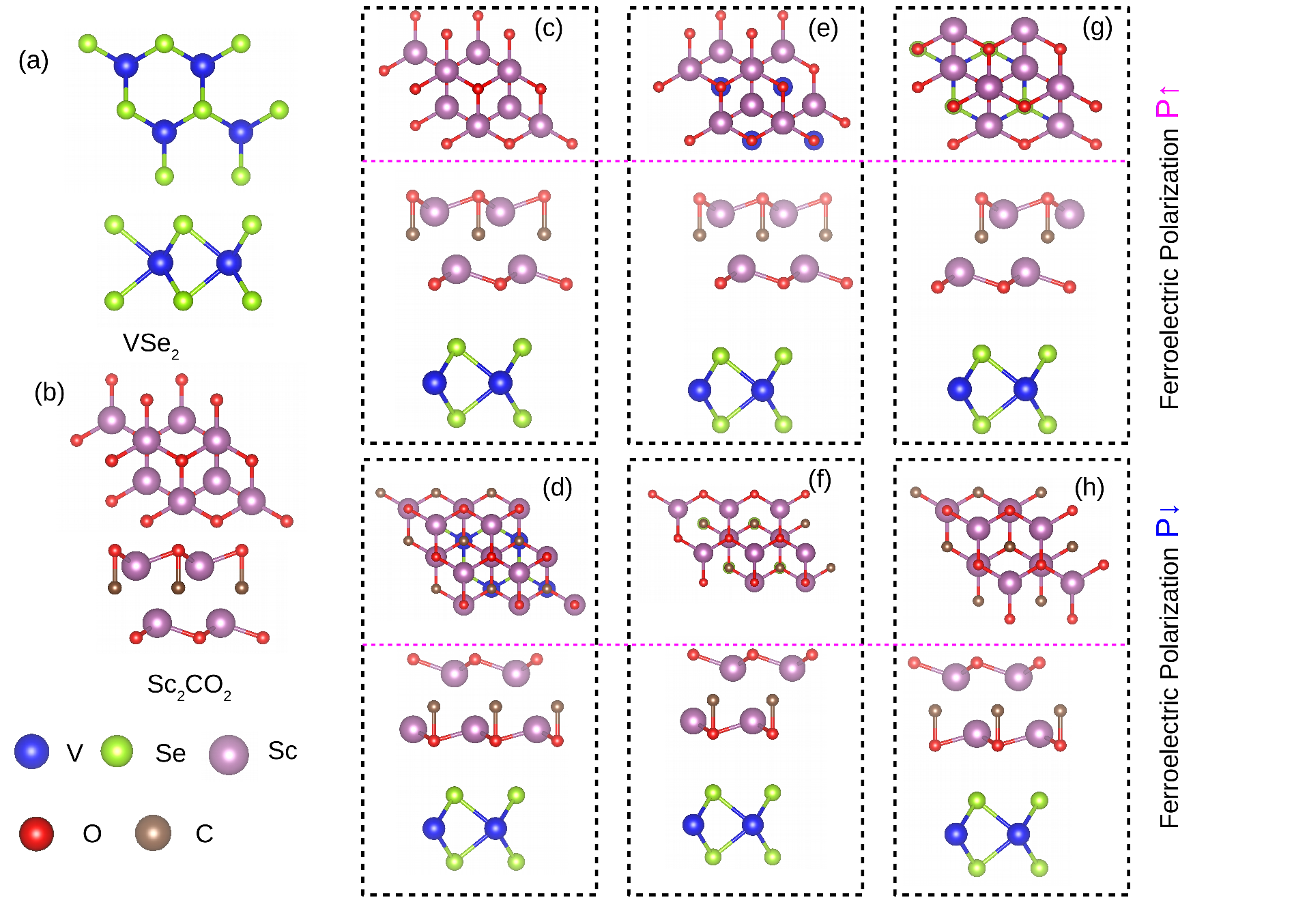}
     \caption{Top and side views of (a)VSe\textsubscript{2} (b) Sc\textsubscript{2}CO\textsubscript{2} monolayer crystal structures. (c),(e),(g) show the stacking patterns of AA-P$\uparrow$, BB-P$\uparrow$ and CC-P$\uparrow$ heterostructures, respectively. (d),(f) and (h) are for the same three stackings when the ferroelectric polarisation is P$\downarrow$. Both top and side views are shown in each case.}
    \label{Fig:1}
\end{figure*}
All calculations in this work are performed using Density Functional Theory (DFT)\cite{dft} as implemented in Vienna ab initio simulation package (VASP) \citep{kresse1996efficient}.  The electron-ion interactions are described using projector augmented wave (PAW) basis set \citep{kresse1999ultrasoft}. The exchange-correlation part of the Hamiltonian is approximated by generalized gradient approximation (GGA) as parameterized by Perdew-Burke-Ernzerhof (PBE) \cite{perdew1996generalized}. The kinetic energy cutoff of the plane wave is set to be 600 eV. The energy and force cutoff for structural optimization are $ 10^{-6} $ eV and 0.01 eV/\AA, respectively. To include the van-der-Waals interactions, the DFT-D3 method is employed \cite{grimme2010consistent}. The Brillouin zone is sampled using Monkhorst-Pack \cite{mp}$ 14\times14\times1$ $k$-point mesh. Dipole corrections are also taken into account in all the calculations. A 20 \AA vacuum is considered along  $z$-direction to avoid any interactions between the adjacent layers.

\section{RESULTS AND DISCUSSIONS}
\subsection{Ferroelectric polarisation driven modifications in the electronic properties of VSe$_{2}$/Sc$_{2}$CO$_{2}$}
\begin{figure*}
    \includegraphics[height=10cm, width=18.00 cm]{Figure_2a.eps}
    \includegraphics[height=10cm, width=18.00 cm]{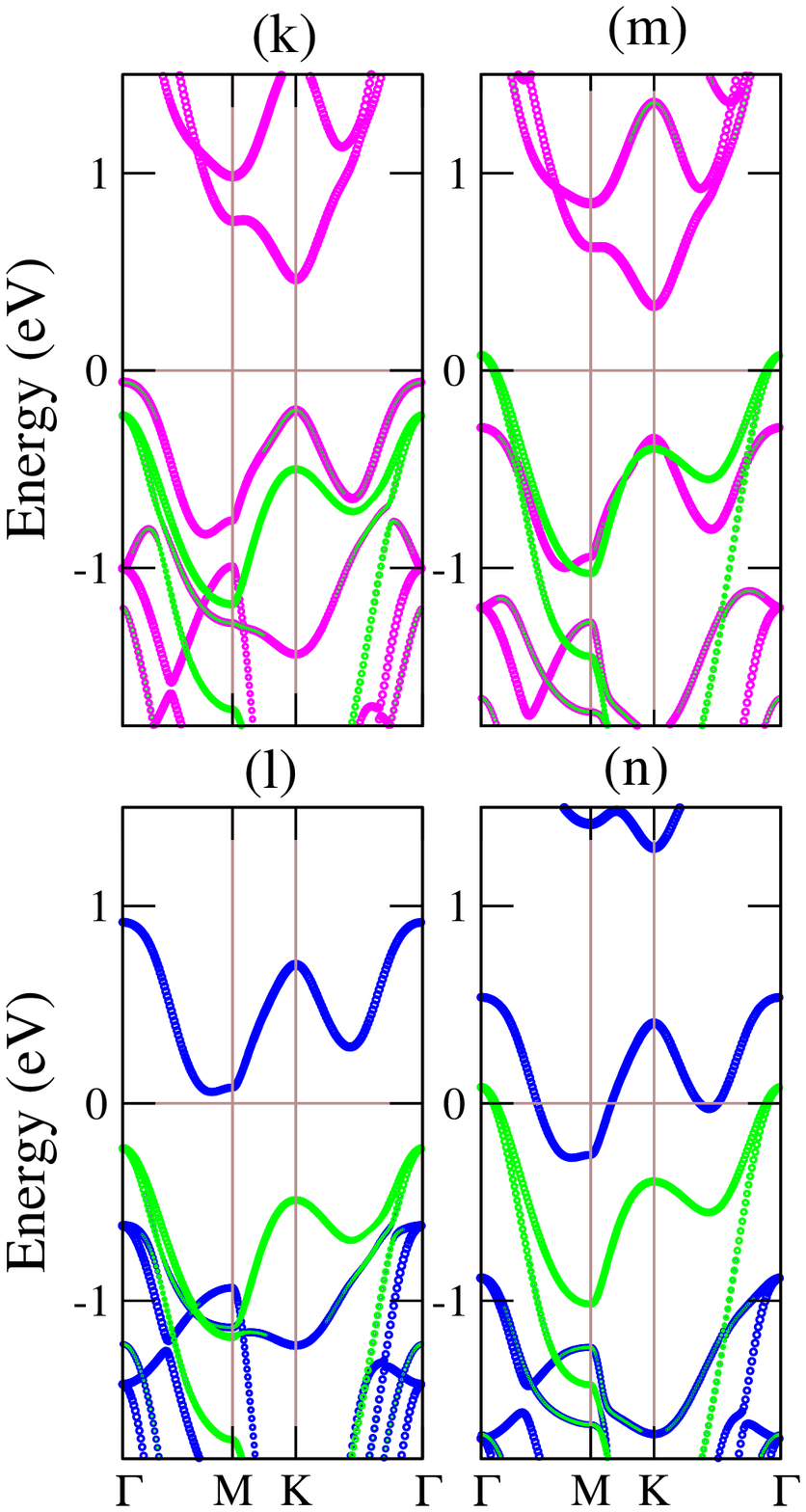}
     \caption{Electronic band structure for (a) Monolayer VSe\textsubscript{2} and (b) Monolayer Sc\textsubscript{2}CO\textsubscript{2}. (c),(d) are the spin-up and spin-down bands of AA-P$\uparrow$ heterostructure, respectively. (e),(f) are the spin-up and spin-down bands of AA-P$\downarrow$ heterostructure, respectively.(g),(h) are the spin-up and spin-down bands of BB-P$\uparrow$ heterostructure, respectively. (i),(j) are the spin-up and spin-down bands of BB-P$\downarrow$ heterostructure,respectively.(k),(l) are the spin-up and spin-down bands of CC-P$\uparrow$ heterostructure, respectively. (m),(n) are the spin-up and spin-down bands of CC-P$\downarrow$ heterostructure,respectively.Green, magenta and blue lines represent contributions from Sc\textsubscript{2}CO\textsubscript{2},spin-up and spin-down states of VSe\textsubscript{2}, respectively.}
    \label{Fig:2}
\end{figure*}

Figure \ref{Fig:1}(a) presents the top and side views of monolayer 2H-VSe$_{2}$, the magnetic component of the heterostructure. It crystallizes in a honeycomb-like structure where each V(Se)-atom is surrounded by six(three) neighboring Se(V) atoms. The V atoms are in a trigonal prismatic crystal field formed by the surrounding ligands. This VSe$_{2}$ phase exhibits ferromagnetism with a very high Curie temperature of 425 K \cite{wang2021ferromagnetism}. Due to its high transition temperature, it can be directly used for applications above room temperature. The lattice constant of VSe\textsubscript{2} from our calculation is 3.33 \AA, which agrees very well with the reported result \cite{fuh2016newtype}. The magnetic moment of each V atom is 1 $\mu B $.

 The ground state structure of MXene monolayer Sc$_{2}$CO$_{2}$, the ferroelectric component of the heterostructure, is shown in Figure \ref{Fig:1}(b). Monolayer Sc$_{2}$CO$_{2}$ possesses out-of-plane ferroelectricity with significantly large electric polarisation \cite{sc2co2abhishek}. The structure lacks inversion symmetry. The O-atom on one of the surfaces occupies the hollow site associated with the Sc atom. The O-atom, on the other surface, occupies the hollow site associated with the C-atom. Consequently, the Sc atom on the lower surface (00-1 surface) is in an octahedral crystal field, while the Sc atom on the top surface (001 surface) is in a trigonal prismatic crystal field. The lattice parameter for Sc\textsubscript{2}CO\textsubscript{2} from our calculations is 3.41 \AA, in excellent agreement with the reported results \cite{cao2022multiferroic,wang2023first}.
 
Our result demonstrates that the lattice mismatch between VSe\textsubscript{2} and Sc\textsubscript{2}CO\textsubscript{2} is nearly 2.4 \% only. Hence, these two monolayers can be vertically stacked to form a 2D van der Waals (vdW) heterostructure. The two ferroelectric polarization states of Sc\textsubscript{2}CO\textsubscript{2} are differentiated by the position of the C atom in the unit cell. When the C atom is close to the upper(lower) surface, the polarisation state is depicted as P$\uparrow$(P$\downarrow$). These two polarisation states can be reversibly switched by applying an electric field and occur through an intermediate antiferroelectric structure \cite{sc2co2abhishek}. The heterostructure with VSe$_{2}$ can be formed with either state of ferroelectric polarisation. Accordingly, we construct six different heterostructures, three each for a given ferroelectric polarisation: (i) AA: The O-atom closest to VSe\textsubscript{2} is directly above the V-atom (Figure \ref{Fig:1}(c),(d)) (ii) BB: The O-atom closest to VSe\textsubscript{2} is directly above the Se atom (Figure \ref{Fig:1}(e),(f))  and (iii) CC: The O-atom closest to VSe\textsubscript{2} is above a V-Se bond in VSe\textsubscript{2} (Figure \ref{Fig:1}(g),{h)).

\begin{figure*}
    \includegraphics[height=12cm, width=18.00 cm]{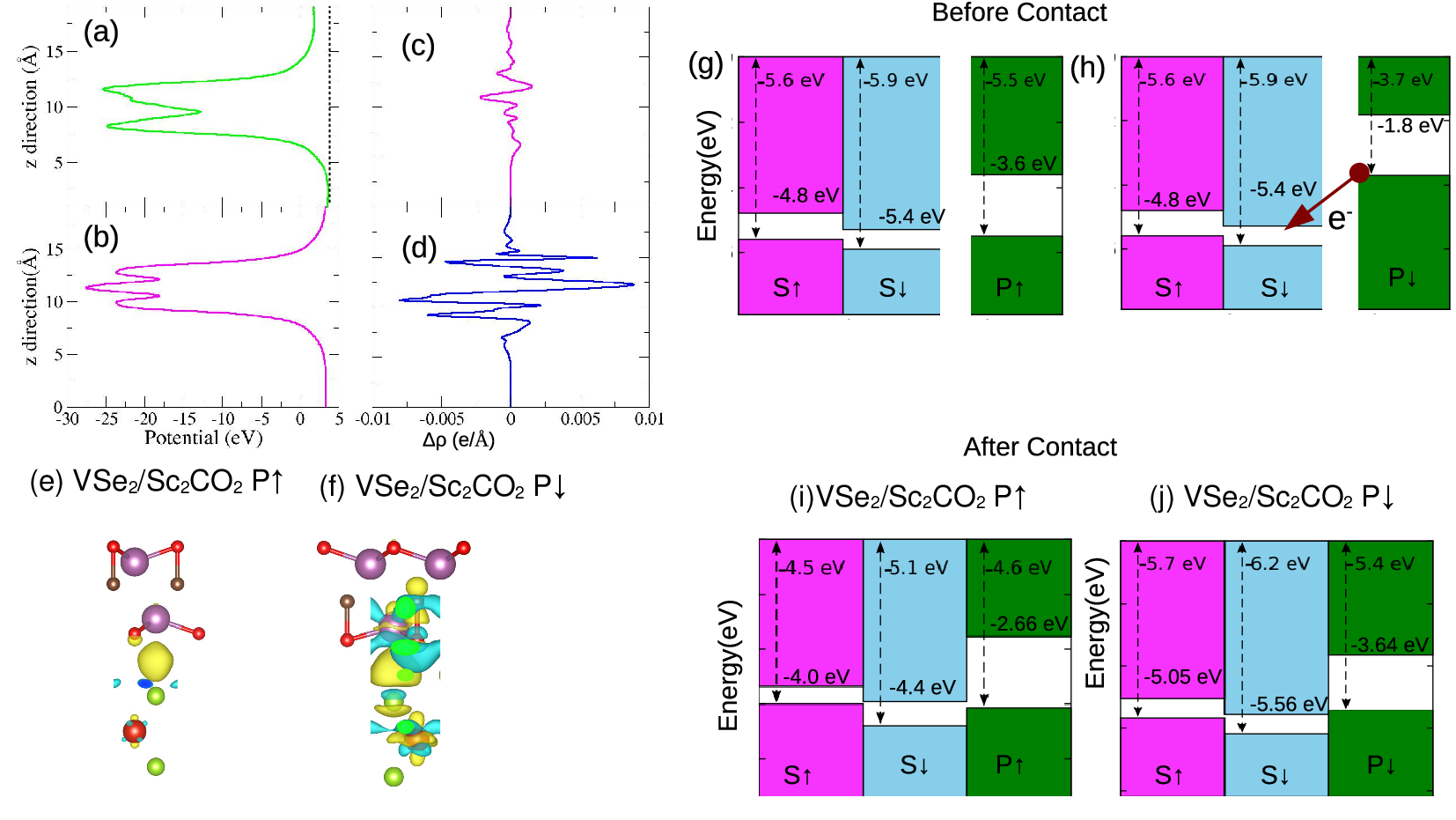}
     \caption{Plane-averaged electrostatic potential of (a)  Sc\textsubscript{2}CO\textsubscript{2}  and (b)VSe\textsubscript{2} along $z$. (c) and (d) are variations in the plane-averaged differential charge density of VSe$_{2}$/Sc$_{2}$CO$_{2}$ (P$\uparrow$)and VSe$_{2}$/Sc$_{2}$CO$_{2}$ (P$\downarrow$)heterostructures along $z$, respectively; (e) and (f) are the corresponding charge density differences. (g) and (h) are the band-alignments of the individual monolayers before contact, while(i) and (j) are the band alignments of VSe$_{2}$/Sc$_{2}$CO$_{2}$(P$\uparrow$)and VSe$_{2}$/Sc$_{2}$CO$_{2}$(P$\downarrow$) heterostructures, respectively. }
    \label{Fig:3}
\end{figure*}

\begin{figure*}
    \includegraphics[height=12cm, width=18.00 cm]{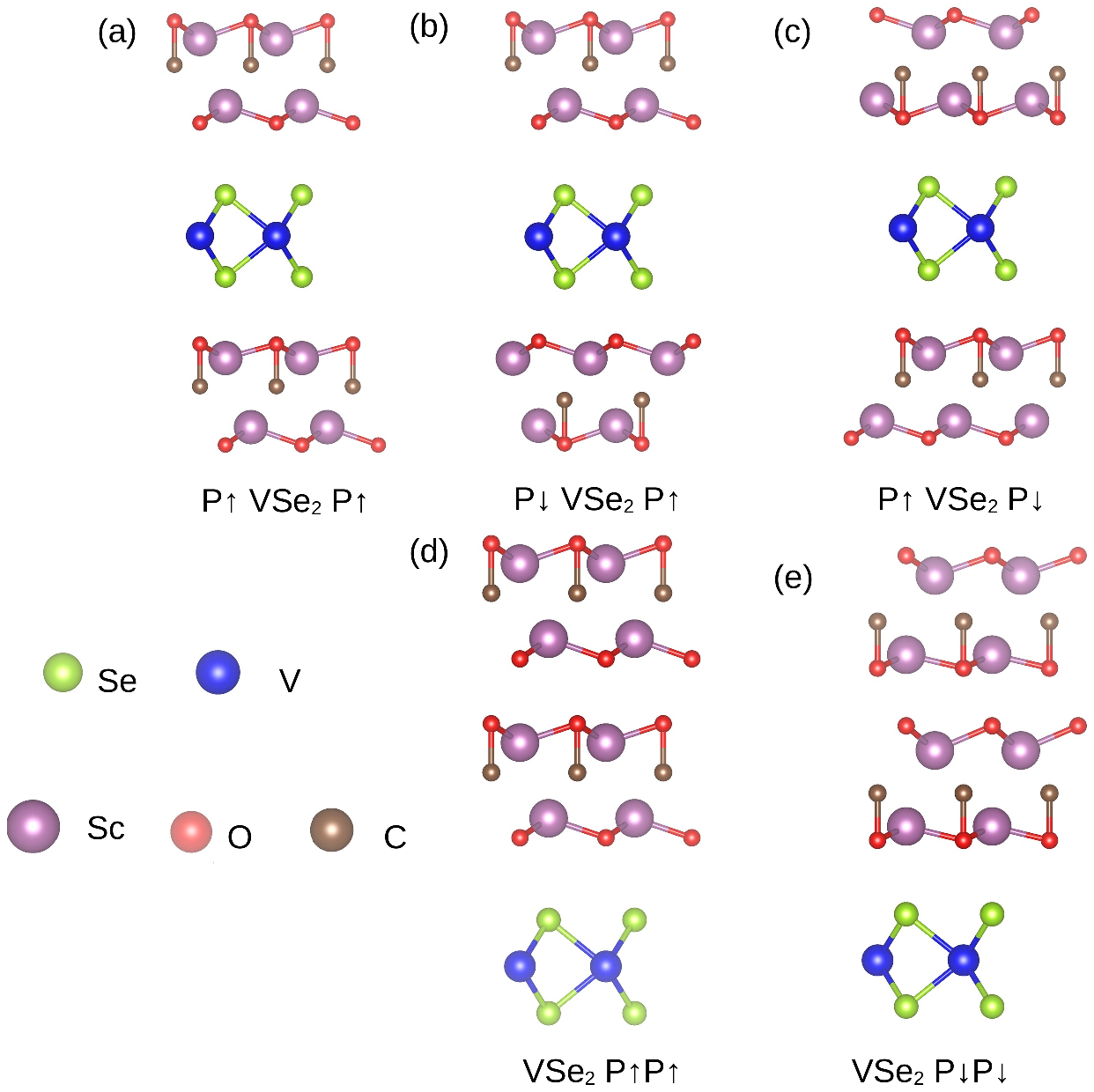}
     \caption{Optimized geometries of  Sc\textsubscript{2}CO\textsubscript{2}/VSe\textsubscript{2}/Sc\textsubscript{2}CO\textsubscript{2} and VSe\textsubscript{2}/Sc\textsubscript{2}CO\textsubscript{2}/Sc\textsubscript{2}CO\textsubscript{2} trilayer heterostructures with differrent polarisations . }
    \label{Fig:4}
\end{figure*}

\begin{table}
\caption{\label{tab:Table1}Calculated parameters of VSe\textsubscript{2}/Sc\textsubscript{2}CO\textsubscript{2} vdw heterostructure . d is interlayer distance, E\textsubscript{b} is interface binding energy.}
\begin{tabular}{ m{0.1\textwidth}m{0.1\textwidth}m{0.14\textwidth}m{0.1\textwidth}}
 \hline
Stacking & d (\AA) & E\textsubscript{b} (meV/\AA\textsuperscript{2}) & Band Gap(eV) \\
\hline
AA-P$\uparrow $ & 2.90 & -19.84 & 0.12 \\
AA-P$\downarrow$ & 2.81 & -31.50 & 0.00\\
BB-P$\uparrow$ & 3.4 & -13.27 & 0.16\\
BB-P$\downarrow$ & 3.35 & -17.5 & 0.00 \\
CC-P$\uparrow$ & 3.02 & -17.8 & 0.12\\
CC-P$\downarrow$ & 2.93 & -31.4 & 0.00\\
\hline
\end{tabular}
\end{table}

The stability of the six heterostructures is assessed by calculating their binding energies given by

$$ E_{b} = (E_{hetero} -E_{VSe_{2}} -E_{Sc_{2}CO_{2}})/A $$

$ E_{hetero} $ is the total energy of the heterostructure. $ E_{VSe_{2}} $ and $ E_{Sc_{2}CO_{2}} $ are the energies of pristine monolayers of VSe\textsubscript{2} and Sc\textsubscript{2}CO\textsubscript{2}, respectively .$ A$ is the surface area of the heterostructure. The results are presented in Table \ref{tab:Table1}. The results imply that the heterostructures can be grown experimentally. The interlayer distance $d$ in each case is also given in Table \ref{tab:Table1}. We find that the interlayer distances decrease when the ferroelectric polarisation of Sc$_{2}$CO$_{2}$ changes from P$\uparrow$ to P$\downarrow$. For a given structural configuration, the maximum changes in the interlayer distances are about 3\% only upon a change in the state of ferroelectric polarisation.

We next investigate the magneto-electric coupling in the bi-layer VSe$_{2}$-Sc$_{2}$CO$_{2}$ by computations of band structures of the heterostructures considered in this work. In Figure \ref{Fig:2}(a) and \ref{Fig:2}(b), we show the band structures of monolayer VSe$_{2}$ and Sc$_{2}$CO$_{2}$, respectively. VSe\textsubscript{2} turns out to be a ferromagnetic semiconductor with a band gap of 0.75 eV(0.53 eV) in the spin-up(spin-down) channel, in agreement with existing results \cite{fuh2016newtype}. Our calculations show Sc\textsubscript{2}CO\textsubscript{2} a semiconductor with an indirect band gap of 1.84 eV, in excellent agreement with other calculations\cite{zhao2019nonvolatile}. The spin-projected band structures of bi-layer VSe\textsubscript{2}/Sc\textsubscript{2}CO\textsubscript{2} for different heterostructures and different ferroelectric polarisations are shown in Figure \ref{Fig:2}(c)-(n). A common feature seen in all cases is that the band structure of the heterostructures is essentially a superposition of bands contributed by the constituents. This is the characteristic of 2D vdW heterostructures with weak coupling between the layers.

The other common feature connected to the magneto-electrical coupling in this system is the dependence of spin projected band structure on the ferroelectric polarisation of MXene Sc$_{2}$CO$_{2}$. In Figure \ref{Fig:2} (c) and (d), we show the spin up and spin down bands of the AA heterostructure, respectively, when the ferroelectric polarisation is P$\uparrow$. We find that the system behaves as a magnetic semiconductor with a band gap of 0.12 eV (Table \ref{tab:Table1}). Here, both the valence band minima (VBM) and conduction band maxima (CBM) are due to contributions from VSe\textsubscript{2}, demonstrating a type-I band alignment. Such heterostructures with type-I band alignment are useful for confining electrons and holes, facilitating carrier recombination, and thus sought after for applications in optical devices. Significantly, the VBM and the CBM are composed of oppositely polarised spin bands. As such, the heterostructure behaves like a Bipolar Magnetic Semiconductor (BMS). In this system, a completely spin polarisation current with reversible spin polarisation can be created and controlled simply by applying a gate voltage \cite{li2012bipolar}. When the polarisation state of  Sc\textsubscript{2}CO\textsubscript{2} ferroelectric is switched to P$\downarrow$, the heterostructure behaves as a half-metal (Figure \ref{Fig:2} (e),(f)). In the spin-up channel, the system remains semiconducting with little changes in the band structure (Figure \ref{Fig:2}(e)). Major changes are observed in the spin-down bands (Figure \ref{Fig:2}(f)). The spin-down band corresponding to VSe$_{2}$, right above Fermi level in case of both monolayer VSe$_{2}$ and the heterostructure with P$\uparrow$ ferroelectric state, is pushed towards lower energy, closing the semiconducting gap, leading to a half-metallic ground state of the heterostructure. 

In experiments, switching between the two states, P$\uparrow$ and P$\downarrow$, can be done by applying an electric field. Sc\textsubscript{2}CO\textsubscript{2}, being ferroelectric, is bistable. As a result, the application of an electric field can change its polarization state. It remains in a given state of polarisation even after removing the electric field, making the switching completely reversible and non-volatile. Figure \ref{Fig:2}, therefore, shows that the ground state of Sc\textsubscript{2}CO\textsubscript{2}/VSe\textsubscript{2} heterostructure can be switched between a magnetic semiconductor and a half-metal by reversible switching of the ferroelectric polarisation. 
\begin{figure*}
    \includegraphics[height=12cm, width=18.00 cm]{Figure_5.eps}
     \caption{(a)-(e) show spin-up band structures for  Sc\textsubscript{2}CO\textsubscript{2}(P$\uparrow$)/ VSe\textsubscript{2}/ Sc\textsubscript{2}CO\textsubscript{2}(P$\uparrow$), Sc\textsubscript{2}CO\textsubscript{2}(P$\downarrow$)/ VSe\textsubscript{2}/ Sc\textsubscript{2}CO\textsubscript{2}(P$\uparrow$) , Sc\textsubscript{2}CO\textsubscript{2}(P$\uparrow$)/ VSe\textsubscript{2}/ Sc\textsubscript{2}CO\textsubscript{2}(P$\downarrow$), Sc\textsubscript{2}CO\textsubscript{2}(P$\uparrow$)/Sc\textsubscript{2}CO\textsubscript{2}(P$\uparrow$)/VSe\textsubscript{2}  and Sc\textsubscript{2}CO\textsubscript{2} (P$\downarrow$)/Sc\textsubscript{2}CO\textsubscript{2}(P$\downarrow$)/VSe\textsubscript{2}, respectively. Corresponding spin-down band structures are shown in (f)-(j), respectively. }
  \label{Fig:5}
\end{figure*} 
\subsection{Coupling Mechanism and interfacial interaction in VSe\textsubscript{2}/Sc\textsubscript{2}CO\textsubscript{2}  heterostructure}
\begin{figure*}
    \includegraphics[height=12cm, width=18.00 cm]{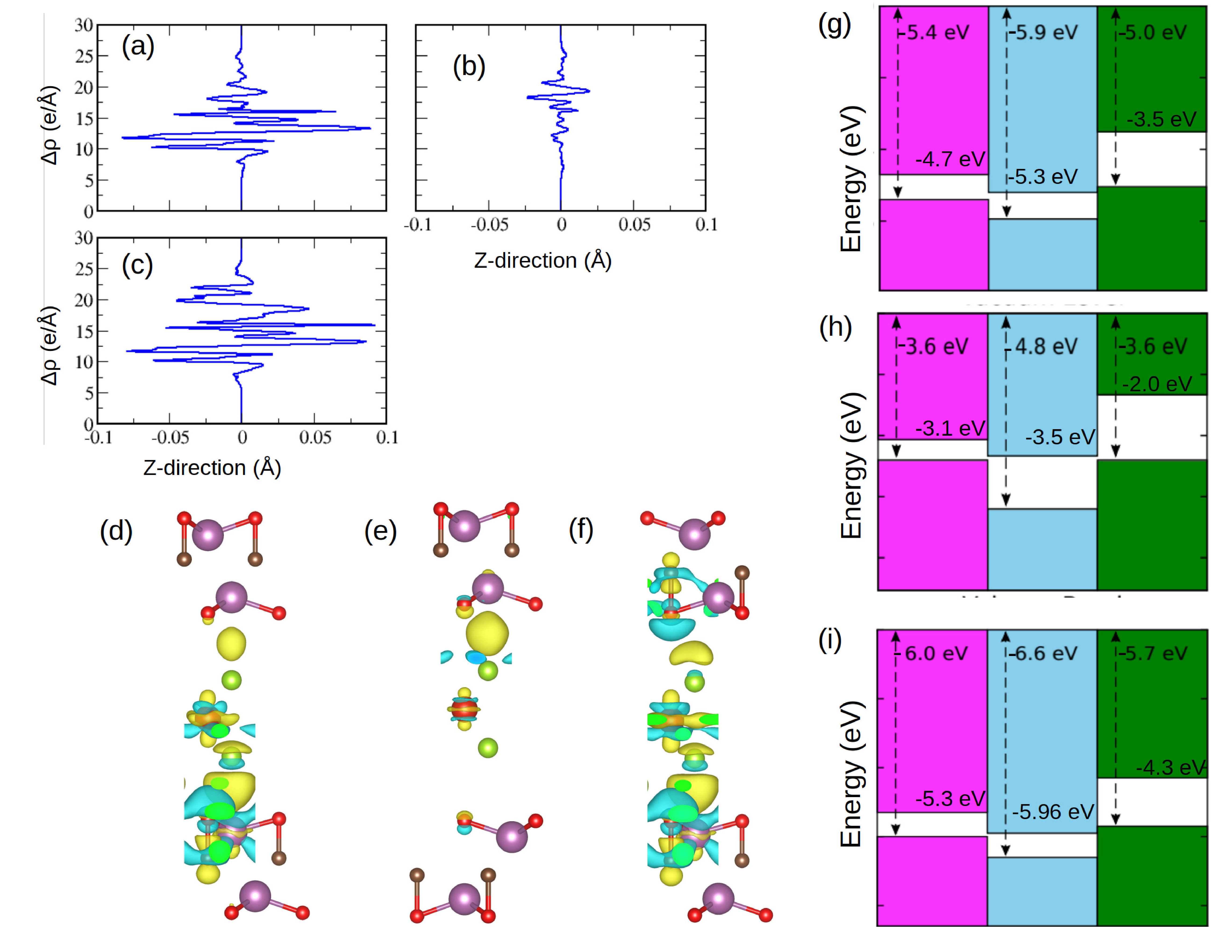}
     \caption{Plane averaged differential charge density (a-c), Charge density differences (d-f) and band alignments (g-i) of  Sc\textsubscript{2}CO\textsubscript{2}(P$\uparrow$)/ VSe\textsubscript{2}/ Sc\textsubscript{2}CO\textsubscript{2}(P$\uparrow$),  Sc\textsubscript{2}CO\textsubscript{2}(P$\downarrow$)/ VSe\textsubscript{2}/ Sc\textsubscript{2}CO\textsubscript{2}(P$\uparrow$)  and  Sc\textsubscript{2}CO\textsubscript{2}(P$\uparrow$)/ VSe\textsubscript{2}/ Sc\textsubscript{2}CO\textsubscript{2}(P$\downarrow$) heterostructures, respectively}
  \label{Fig:6}
\end{figure*}  
The observed changes in the electronic structures of VSe$_{2}$/Sc$_{2}$CO$_{2}$ upon changing the ferroelectric polarisation state can be understood by analyzing the changes in the charge distribution or coupling across the monolayer interfaces. Due to broken spatial symmetry, the work function associated with two surfaces of Sc\textsubscript{2}CO\textsubscript{2} is different by about 1.8 $ eV $ as seen from the plane-averaged potential versus distance in the $z$-direction plot (Figure \ref{Fig:3}(a)). No such asymmetry across the surfaces is observed in the case of VSe$_{2}$ (Figure \ref{Fig:3}(b)). The calculated Work functions of VSe$_{2}$ monolayer, the C$_{top}$ side and C$_{bottom}$ side of Sc$_{2}$CO$_{2}$ monolayer are -5.46 eV, -4.96 eV and -3.13 eV, respectively; C$_{top}$(C$_{bottom}$) refers to the position of C-atom layer in Sc$_{2}$CO$_{2}$ when the ferroelectric polarisation is P$\uparrow$(P$\downarrow$). Thus, when VSe\textsubscript{2} is stacked with Sc\textsubscript{2}CO\textsubscript{2}, depending on which surface of Sc\textsubscript{2}CO\textsubscript{2} is used for stacking, charge transfer across the interface differs significantly. 
When Sc\textsubscript{2}CO\textsubscript{2} is in the P$\uparrow$ polarisation state, the spin-up(S$\uparrow$) and spin-down(S$\downarrow$) valence band maxima (VBM) (conduction band minima(CBM)) of VSe\textsubscript{2} are lower(higher) than the conduction band minima (CBM)(valence band maxima (VBM)) of Sc\textsubscript{2}CO\textsubscript{2} (Figure\ref{Fig:3}(g)). Thus, in VSe$_{2}$/Sc$_{2}$CO$_{2}$(P$\uparrow$) heterostructure, there is no electron transfer between Sc\textsubscript{2}CO\textsubscript{2} and VSe\textsubscript{2}. Nevertheless, due to differences in the work functions of Sc\textsubscript{2}CO\textsubscript{2} and VSe\textsubscript{2} the interfacial interaction causes a spontaneous flow of electrons from Sc\textsubscript{2}CO\textsubscript{2} to VSe\textsubscript{2}. Accordingly, the Fermi level of Sc\textsubscript{2}CO\textsubscript{2} shifts slightly towards lower energy, and that of VSe\textsubscript{2} shifts slightly towards higher energy, reaching a common level. As a result, a small amount of electron accumulates in VSe\textsubscript{2}, giving rise to an internal electric field. The charge transfer, however, is not sufficient to close the energy gap, resulting in a type I band alignment (Figure \ref{Fig:3}(i)). The variations in the plane-averaged differential charge density with distance along $z$ (Figure \ref{Fig:3}(c)) and charge density difference (Figure \ref{Fig:3}(e)) results corroborate this. 

When the polarization of Sc\textsubscript{2}CO\textsubscript{2} changes to P$\downarrow$ (Figure \ref{Fig:3}(h)),the VBM of Sc\textsubscript{2}CO\textsubscript{2}is at a higher position than CBM of VSe\textsubscript{2} associated with either spin band. Therefore, electron transfer can occur from the valence band of Sc\textsubscript{2}CO\textsubscript{2} to the conduction band of the spin-down channel of VSe\textsubscript{2}. Due to this, the spin-down conduction band of VSe$_{2}$ is shifted towards lower energy considerably, closing the semiconducting gap in this spin channel and giving rise to metallic behavior (Figure \ref{Fig:3}(j)). The band alignment shown in Figure \ref{Fig:3}(j) indicates that this heterostructure exhibits an almost broken-gap type-III alignment. Heterostructures with such band alignment can facilitate tunneling of electrons directly from the VBM of Sc\textsubscript{2}CO\textsubscript{2} to the CBM of VSe\textsubscript{2}. Therefore, band-to-band tunneling (BTBT) is possible, making these heterostructures highly desirable for tunnel field-effect transistors. The plane-averaged differential charge density (Figure \ref{Fig:3}(d)) and charge density difference(Figure \ref{Fig:3}(f)) plots show a significant amount of charge transfer across the interface justifying the band alignment described above. Such large charge transfer across the interface can be explained in terms of the significant difference in the Work functions of VSe$_{2}$ and C$_{bottom}$ side of the Sc$_{2}$CO$_{2}$ monolayer.  
\subsection{Ferroelectric polarisation driven modifications in the electronic properties of Sc\textsubscript{2}CO\textsubscript{2}/VSe\textsubscript{2}/Sc\textsubscript{2}CO\textsubscript{2}} 
\begin{table*}
\caption{\label{tab:Table2}Calculated interface binding energies (E$_{b}$)of trilayer vdw heterostructures (in meV/\AA$^{2}$).}
\begin{tabular}{ m{0.40\textwidth}m{0.1\textwidth}}
 \hline
System &  E\textsubscript{b}\\
\hline
Sc\textsubscript{2}CO\textsubscript{2}(P$\uparrow$)/VSe\textsubscript{2}/Sc\textsubscript{2}CO\textsubscript{2}(P$\uparrow$) &  -30.00\\
Sc\textsubscript{2}CO\textsubscript{2}(P$\downarrow$)/VSe\textsubscript{2}/Sc\textsubscript{2}CO\textsubscript{2}(P$\uparrow$) &  -21.75  \\
Sc\textsubscript{2}CO\textsubscript{2}(P$\uparrow$)/VSe\textsubscript{2}/Sc\textsubscript{2}CO\textsubscript{2}(P$\downarrow$) &  -31.24  \\
VSe\textsubscript{2}/Sc\textsubscript{2}CO\textsubscript{2}(P$\uparrow$)Sc\textsubscript{2}CO\textsubscript{2}(P$\uparrow$) &  -29.60  \\
VSe\textsubscript{2}/Sc\textsubscript{2}CO\textsubscript{2}(P$\downarrow$)Sc\textsubscript{2}CO\textsubscript{2}(P$\downarrow$) &  -51.95  \\
\hline
\end{tabular}
\end{table*}    
Next, we investigate the interrelations between a number of layers in the VSe$_{2}$-Sc$_2$CO$_2$ heterostructure, the ferroelectric polarisation, and the ground state electronic structure. For this purpose, we construct a tri-layer Sc\textsubscript{2}CO\textsubscript{2}/VSe\textsubscript{2}/Sc\textsubscript{2}CO\textsubscript{2} heterostructure. Since the AA stacking configuration has the lowest binding energy in the bi-layer heterostructure, and since a different stacking configuration did not produce any qualitatively different result, we have considered only the AA stacking configuration for this trilayer heterostructure. Three different arrangements, (a) Sc\textsubscript{2}CO\textsubscript{2}(P$\uparrow$)/ VSe\textsubscript{2}/ Sc\textsubscript{2}CO\textsubscript{2}(P$\uparrow$) (b)Sc\textsubscript{2}CO\textsubscript{2}(P$\downarrow$)/ VSe\textsubscript{2}/ Sc\textsubscript{2}CO\textsubscript{2}(P$\uparrow$) (c) Sc\textsubscript{2}CO\textsubscript{2}(P$\uparrow$)/ VSe\textsubscript{2}/ Sc\textsubscript{2}CO\textsubscript{2}(P$\downarrow$), are considered. Additionally,two more arrangements, Sc\textsubscript{2}CO\textsubscript{2}(P$\uparrow$)/Sc\textsubscript{2}CO\textsubscript{2}(P$\uparrow$)/VSe\textsubscript{2}  and Sc\textsubscript{2}CO\textsubscript{2}(P$\downarrow$)/Sc\textsubscript{2}CO\textsubscript{2}(P$\downarrow$)/VSe\textsubscript{2} are considered for the purpose of comparison. The arrangements are shown in Figure \ref{Fig:4}. Binding energies corresponding to them are given in Table \ref{tab:Table2}. The negative values of binding energies indicate that the heterostructures can form with these arrangements.\\

Figure \ref{Fig:5} shows the spin-projected band structures of these trilayer arrangements of VSe$_{2}$-Sc$_{2}$CO$_{2}$ heterostructure.Among the three arrangements with VSe$_{2}$ as the intermediate layer, half-metallic ground state is observed for Sc$_{2}$CO$_{2}$(P$\uparrow$)/VSe$_{2}$/Sc$_{2}$CO$_{2}$(P$\uparrow$) (Figure \ref{Fig:5}(a),(f)) and Sc$_{2}$CO$_{2}$(P$\uparrow$)/VSe$_{2}$/Sc$_{2}$CO$_{2}$(P$\downarrow$)(Figure \ref{Fig:5}(c),(h)). In both cases, the spin-down(spin-up) band is metallic (semiconducting). In contrast, Sc\textsubscript{2}CO\textsubscript{2}(P$\downarrow$)/ VSe\textsubscript{2}/ Sc\textsubscript{2}CO\textsubscript{2}(P$\uparrow$) heterostructure is semi-conducting in both spin channels (Figure \ref{Fig:5}(b),(g)). Thus, semiconductor to half-metallic ground state can be realized in these tri-layer heterostructures by changing polarisations of the MXene layers through the application of an external electric field. This implies that the tri-layer heterostructures of Sc$_{2}$CO$_{2}$-VSe$_{2}$ with the latter sandwiched between two MXene layers can act as a multiferroic. This is not so with the other two tri-layers, which basically are heterostructures of a bi-layer Sc$_{2}$CO$_{2}$ and a mono-layer of VSe$_{2}$.Upon reversal of polarisation in the bi-layer Sc$_{2}$CO$_{2}$ from P$\uparrow$ to P$\downarrow$, the heterostructure ground state changes from a semi-metallic to a metallic one only(Figures 5(d),(i) and 5(e),(j) ).
\subsection{Coupling Mechanism and interfacial interaction in Sc\textsubscript{2}CO\textsubscript{2}/VSe\textsubscript{2}/Sc\textsubscript{2}CO\textsubscript{2} heterostructure}
\begin{figure*}
    \includegraphics[height=10cm, width=14.00 cm]{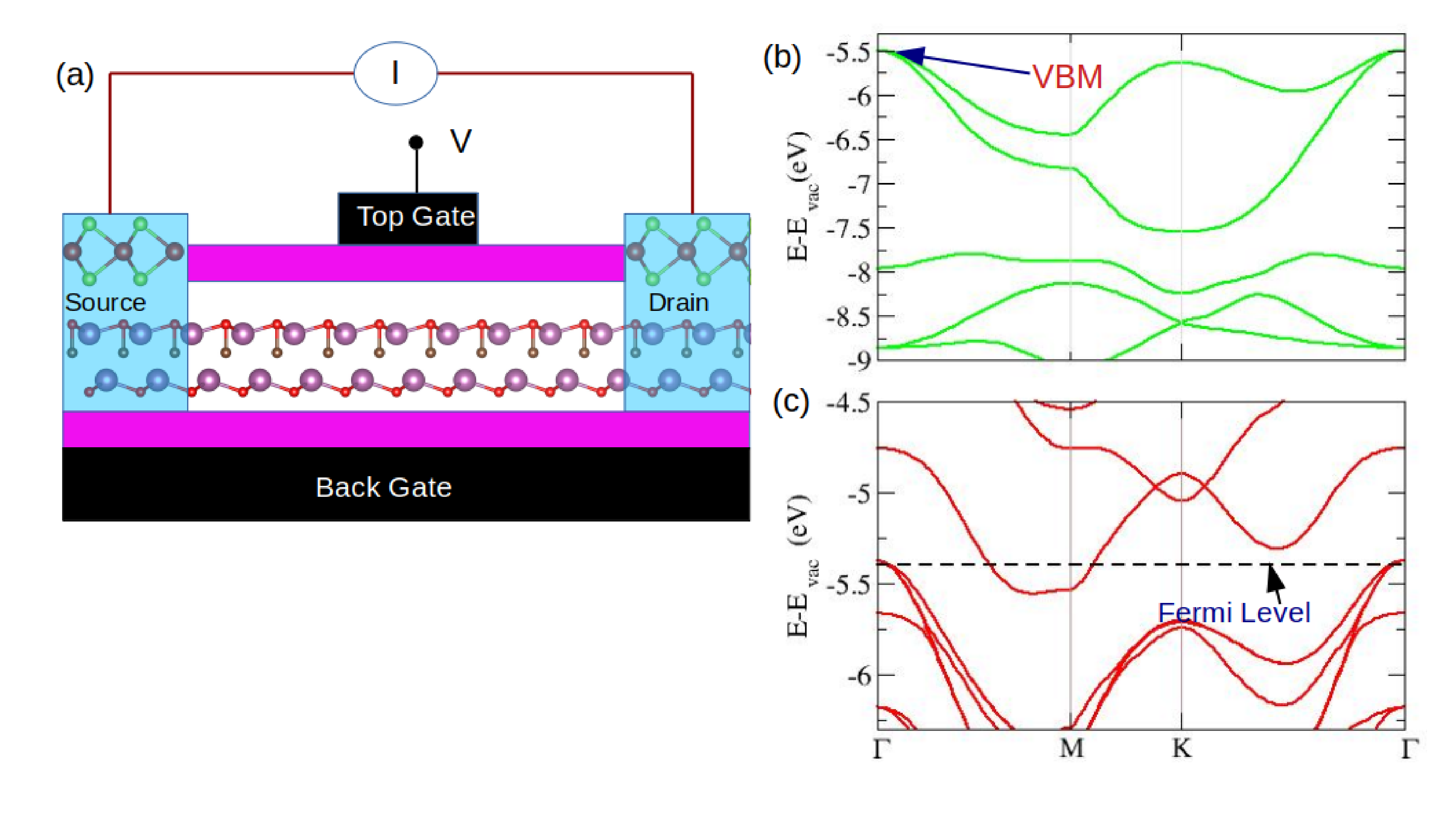}
     \caption{(a) Schematic diagram of VSe\textsubscript{2}-contact Sc\textsubscript{2}CO\textsubscript{2} FET (b) VBM of semiconducting Sc\textsubscript{2}CO\textsubscript{2}  (c) Fermi level of metal electrode  Sc\textsubscript{2}CO\textsubscript{2}(P$\downarrow$)/VSe\textsubscript{2}  }
  \label{Fig:7}
\end{figure*}  
As observed in the case of the bilayer heterostructures, the changes in the electronic structures with stacking and/or electric polarization can be understood through an analysis of charge transfers across the interfaces. The stacking of layers in Sc$_{2}$CO$_{2}$(P$\uparrow$)/VSe$_{2}$/Sc$_{2}$CO$_{2}$(P$\uparrow$) (Figure \ref{Fig:4}(a)) resembles a combination of two bi-layers VSe$_{2}$-Sc$_{2}$CO$_{2}$(P$\uparrow$) (Figure \ref{Fig:1}(g)) and VSe$_{2}$-Sc$_{2}$CO$_{2}$(P$\downarrow$) (Figure \ref{Fig:1}(h)). The top (bottom) surface of the tri-layer has the C$_{bottom}$(C$_{top}$) side of MXene Sc$_{2}$CO$_{2}$ interfacing with VSe$_{2}$. Since the differences in the Work functions of C$_{top}$(C$_{bottom}$) side of Sc$_{2}$CO$_{2}$ and VSe$_{2}$ is small(large), the charge transfer across the top (bottom)interface of the tri-layer heterostructure should be negligible (significant) as was the cases for the bi-layer heterostructures. 
The variations in the plane averaged differential charge density with $z$ (Figure \ref{Fig:6}(a)) and the charge density differences (Figure \ref{Fig:6}(d)) demonstrate this. Due to such charge transfer, the spin-down conduction band of VSe\textsubscript{2} shifts downwards in the energy, closing the semiconducting gap and resulting in a metallic behavior in this spin channel. This makes the ground state of this tri-layer heterostructure half-metallic. The alignment of bands in this heterostructure, shown in Figure \ref{Fig:6}(g), indicates a broken gap type-III band alignment.

In contrast,in Sc$_{2}$CO$_{2}$(P$\downarrow$)/VSe$_{2}$/Sc$_{2}$CO$_{2}$(P$\uparrow$) both interfaces consist of C$_{bottom}$ part of Sc$_{2}$CO$_{2}$(Figure \ref{Fig:4}(b)). The expectation, therefore, is that apart from the small charge transfer due to the internal electric field, there will be no significant charge transfer across the interfaces. Consequently, the Sc\textsubscript{2}CO\textsubscript{2}(P$\downarrow$)/VSe\textsubscript{2}/Sc\textsubscript{2}CO\textsubscript{2}(P$\uparrow$) heterostructure will have a gap opening in both spin channels making the system semiconducting. This explains the band structure corresponding to this heterostructure (Figure \ref{Fig:5}(b)). The variations in the planar averaged differential charge density (Figure \ref{Fig:6}(b)) and the charge density differences (Figure \ref{Fig:6}(e)) support this argument. The band alignment diagram of this heterostructure, shown in Figure \ref{Fig:6}(h), indicates a band alignment of Type-I.

In the Sc\textsubscript{2}CO\textsubscript{2}(P$\uparrow$)/VSe\textsubscript{2}/Sc\textsubscript{2}CO\textsubscript{2}(P$\downarrow$) configuration shown in Figure \ref{Fig:4}(c), both interfaces consist of C$_{top}$ side of Sc$_{2}$CO$_{2}$.As a result, significant charge transfer occurs across both interfaces(Figures \ref{Fig:6}(c),(f)), leading to the closing of the semiconducting gap in the spin-down channel, making the ground state half-metallic. Consequently, the band alignment (Figure \ref{Fig:6}(i)) is a broken Type-III one. The results imply that by applying an electric field, one can switch between a type-I and a broken-gap Type-III system.
\section{Applications}
\subsection{Field-effect transistor(FET)}
\begin{figure*}
    \includegraphics[height=12cm, width=14.00 cm]{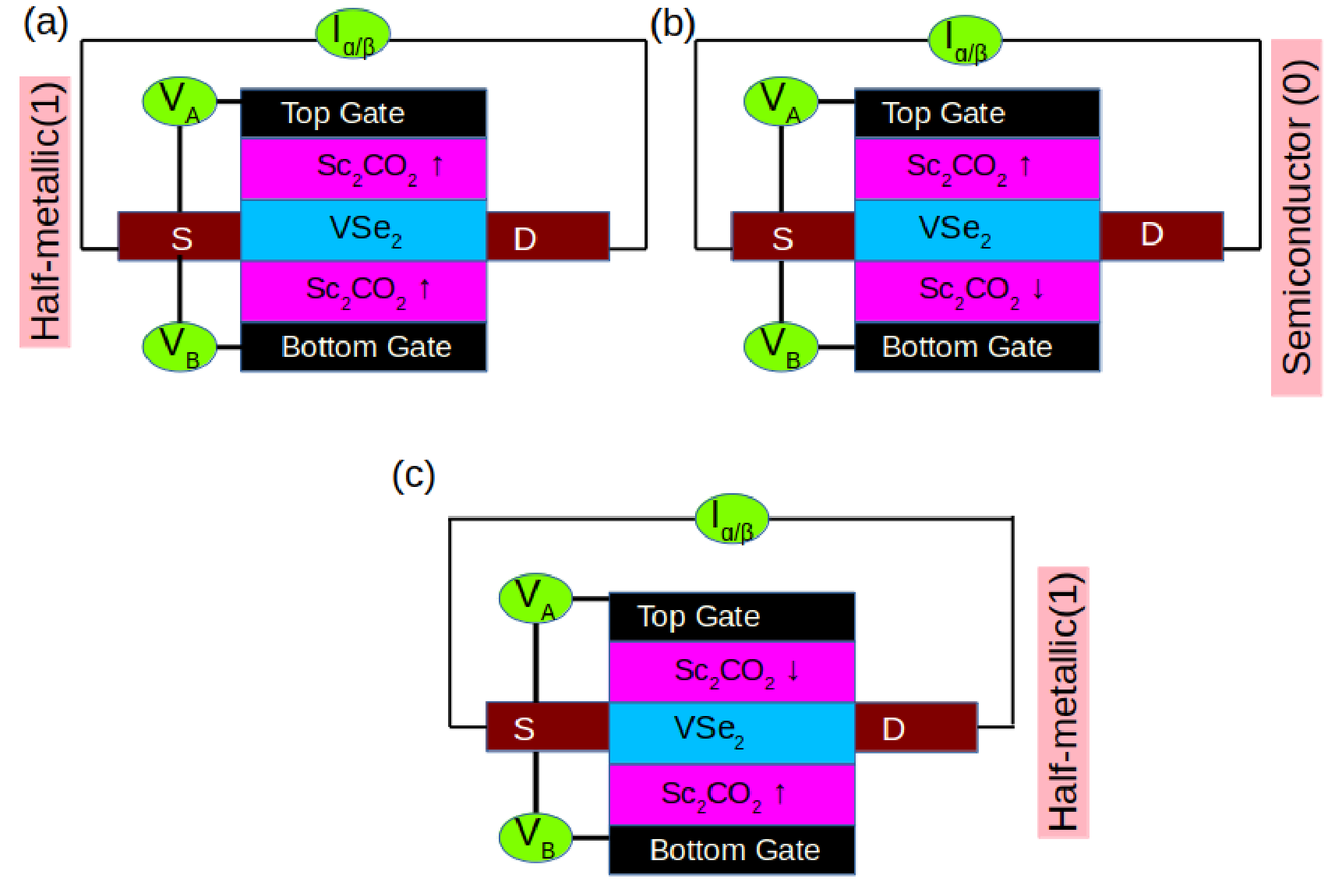}
     \caption{Schematic diagram of atom-thick multiferroic memory device using  Sc\textsubscript{2}CO\textsubscript{2}/VSe\textsubscript{2}/Sc\textsubscript{2}CO\textsubscript{2} heterostructure.}
  \label{Fig:8}
\end{figure*}  
Since VSe\textsubscript{2} becomes half-metallic when interfaced with Sc\textsubscript{2}CO\textsubscript{2}(P$\downarrow$), it can form a van der Waals-contacted metal-semiconductor junction (MSJ) with Sc\textsubscript{2}CO\textsubscript{2}. Such MSJ can play a significant role in FET as the performance of the FET depends strongly on the height of the Schottky barrier(SB) across the junction\cite{zhao2019nonvolatile} since this barrier provides resistance to charge carrier injection across the junction, the smaller the barrier, the lower the resistance, resulting in better FET performance. The SB height for hole injection can be estimated as $ \phi_{h} = E_{F}-E_{VBM} $. For electron injection, it is $ \phi_{e} = E_{CBM}-E_{F}$; $E_{F}$ is the Fermi energy of the heterostructure while $E_{CBM}$ and $E_{VBM}$ are the conduction band minimum and the valence band maximum of semiconducting Sc\textsubscript{2}CO\textsubscript{2}. The schematic diagram for a field-effect transistor (FET) using this heterostructure is shown in Figure \ref{Fig:7}(a) while the position of the VBM of Sc$_{2}$CO$_{2}$ and Fermi level of the heterostructure is shown in Figure \ref{Fig:7}(b) and (c), respectively. The fermi level is slightly higher than VBM for the system under consideration. Consequently, $\phi_{h}$ is only 0.09 eV. This implies that the MSJ constructed out of this system will have a very small SB for hole injection. Further, for better performance of the FET, the semiconductor used in the heterostructure should also have a moderate bandgap and high carrier mobility. Sc\textsubscript{2}CO\textsubscript{2}, with a bandgap of 1.83 eV, is a moderate bandgap semiconductor. It is found to possess high hole mobility as well \cite{dong2019magnetoelectricity}. Thus, half-metallic VSe$_{2}$-contacted Sc$_{2}$CO$_{2}$ FET will have a very small SB and a high on/off ratio.
\subsection{Prototype of a multiferroic memory}
 Two-dimensional (2D) ferroelectrics and 2D ferromagnets are promising nanodevices for information storage. While 2D ferromagnets suffer from non-volatility, 2D ferroelectrics are highly suitable for writing data but are less efficient for reading, the latter operation often being destructive. Memory devices constructed from Sc\textsubscript{2}CO\textsubscript{2}/VSe\textsubscript{2} or Sc\textsubscript{2}CO\textsubscript{2}/VSe\textsubscript{2}/Sc\textsubscript{2}CO\textsubscript{2} can overcome this difficulty. Non-volatile nanodevices for data storage can be constructed out of these heterostructures by utilizing the magnetoelectric coupling. In this nanodevice, the mechanism of data writing is the same as that of isolated monolayer Sc\textsubscript{2}CO\textsubscript{2}, achieved by switching the polarization state through the application of an electric field. Thus, isolated ferroelectrics' efficient data writing properties are retained in these multiferroics.
The data reading part is done the following way: the multiferroic heterostructure (both bilayer or trilayer structure) is a half-metal or semiconductor, depending on the polarization state of Sc\textsubscript{2}CO\textsubscript{2}. As a result, the electric signal of the heterostructure will differ from state to state. The destructive effect observed in monolayer ferroelectrics can be genuinely avoided by reading the magnetoelectric coupling-induced signal differences with a damage-free current \cite{cao2022multiferroic}. A prototype of this device using trilayer Sc$_{2}$CO$_{2}$/VSe$_{2}$/Sc$_{2}$CO$_{2}$ is shown in Figure \ref{Fig:8}. Here, the polarisation states of the two Sc\textsubscript{2}CO\textsubscript{2} layers can be controlled by the application of two different voltages. Voltage V\textsubscript{A}, applied between the top gate and VSe\textsubscript{2}, helps to control the polarisation state of the top Sc\textsubscript{2}CO\textsubscript{2} layer. Similarly, V\textsubscript{B}, applied between the bottom gate and VSe\textsubscript{2}, helps to control the polarisation state of the bottom Sc\textsubscript{2}CO\textsubscript{2} layer.
\section{Conclusion}
Using first-principles-based DFT calculations, we have demonstrated that the electronic properties of bi-layer VSe\textsubscript{2}/Sc\textsubscript{2}CO\textsubscript{2} and tri-layer Sc\textsubscript{2}CO\textsubscript{2}/VSe\textsubscript{2}/Sc\textsubscript{2}CO\textsubscript{2} heterostructures can be reversibly tuned through the application of an electric field. Depending on the polarization states of Sc\textsubscript{2}CO\textsubscript{2}, these heterostructures exhibit semiconducting or half-metallic behavior. The switching mechanism can be understood based on the charge redistribution across the heterostructure interface. The heterostructures exhibit broken-gap type-III band alignment in their half-metallic state, making them ideal for tunnel field-effect transistors (TFETs). Further,
the magneto-electric coupling in these heterostructures leading to the electric-field controlled switching of the electronic state provides an ideal platform for a non-volatile atom-thick high-density data storage. Since a number of MXene Ferroelectrics with substantial polarization have been predicted from DFT-based high throughput calculations very recently \cite{mxenefe}, this work paves the ground for further exploration of MXene-TMDC heterostructures of different thicknesses towards the realization of more memory devices at the nanoscale. 

\begin{mcitethebibliography}{49}
	\providecommand*{\natexlab}[1]{#1}
	\providecommand*{\mciteSetBstSublistMode}[1]{}
	\providecommand*{\mciteSetBstMaxWidthForm}[2]{}
	\providecommand*{\mciteBstWouldAddEndPuncttrue}
	{\def\EndOfBibitem{\unskip.}}
	\providecommand*{\mciteBstWouldAddEndPunctfalse}
	{\let\EndOfBibitem\relax}
	\providecommand*{\mciteSetBstMidEndSepPunct}[3]{}
	\providecommand*{\mciteSetBstSublistLabelBeginEnd}[3]{}
	\providecommand*{\EndOfBibitem}{}
	\mciteSetBstSublistMode{f}
	\mciteSetBstMaxWidthForm{subitem}
	{(\emph{\alph{mcitesubitemcount}})}
	\mciteSetBstSublistLabelBeginEnd{\mcitemaxwidthsubitemform\space}
	{\relax}{\relax}
	
	\bibitem[Yakout(2020)]{spintronics}
	S.~M. Yakout, \emph{Journal of Superconductivity and Novel Magnetism}, 2020,
	\textbf{33}, 2557\relax
	\mciteBstWouldAddEndPuncttrue
	\mciteSetBstMidEndSepPunct{\mcitedefaultmidpunct}
	{\mcitedefaultendpunct}{\mcitedefaultseppunct}\relax
	\EndOfBibitem
	\bibitem[Sarma(2001)]{sdsarma}
	S.~D. Sarma, \emph{American Scientist}, 2001, \textbf{89}, 516\relax
	\mciteBstWouldAddEndPuncttrue
	\mciteSetBstMidEndSepPunct{\mcitedefaultmidpunct}
	{\mcitedefaultendpunct}{\mcitedefaultseppunct}\relax
	\EndOfBibitem
	\bibitem[Tian \emph{et~al.}(2012)Tian, Bakaul, and Wu]{oxidenano}
	Y.~Tian, S.~R. Bakaul and T.~Wu, \emph{Nanoscale}, 2012, \textbf{4}, 1529\relax
	\mciteBstWouldAddEndPuncttrue
	\mciteSetBstMidEndSepPunct{\mcitedefaultmidpunct}
	{\mcitedefaultendpunct}{\mcitedefaultseppunct}\relax
	\EndOfBibitem
	\bibitem[El-Ghazaly \emph{et~al.}(2020)El-Ghazaly, Gorchon, Wilson, Pattabi,
	and Bokor]{ultraspin}
	A.~El-Ghazaly, J.~Gorchon, R.~B. Wilson, A.~Pattabi and J.~Bokor, \emph{Journal
		of Magnetism and Magnetic Materials}, 2020, \textbf{502}, 166478\relax
	\mciteBstWouldAddEndPuncttrue
	\mciteSetBstMidEndSepPunct{\mcitedefaultmidpunct}
	{\mcitedefaultendpunct}{\mcitedefaultseppunct}\relax
	\EndOfBibitem
	\bibitem[Lu \emph{et~al.}(2016)Lu, Chen, Kabir, Stan, and Wolf]{spintronics1}
	J.~W. Lu, E.~Chen, M.~Kabir, M.~R. Stan and S.~A. Wolf, \emph{International
		Materials Review}, 2016, \textbf{61}, 456\relax
	\mciteBstWouldAddEndPuncttrue
	\mciteSetBstMidEndSepPunct{\mcitedefaultmidpunct}
	{\mcitedefaultendpunct}{\mcitedefaultseppunct}\relax
	\EndOfBibitem
	\bibitem[Joshi(2016)]{spintronics2}
	V.~K. Joshi, \emph{Engineering Science and technology, an International
		Journal}, 2016, \textbf{19}, 1503\relax
	\mciteBstWouldAddEndPuncttrue
	\mciteSetBstMidEndSepPunct{\mcitedefaultmidpunct}
	{\mcitedefaultendpunct}{\mcitedefaultseppunct}\relax
	\EndOfBibitem
	\bibitem[Gurney(2008)]{spintronics3}
	B.~A. Gurney, \emph{AAPPS Bulletin}, 2008, \textbf{18}, 18\relax
	\mciteBstWouldAddEndPuncttrue
	\mciteSetBstMidEndSepPunct{\mcitedefaultmidpunct}
	{\mcitedefaultendpunct}{\mcitedefaultseppunct}\relax
	\EndOfBibitem
	\bibitem[Fert(2016)]{spintronics4}
	A.~Fert, \emph{Angewandte Chemie International Edition}, 2016, \textbf{19},
	1503\relax
	\mciteBstWouldAddEndPuncttrue
	\mciteSetBstMidEndSepPunct{\mcitedefaultmidpunct}
	{\mcitedefaultendpunct}{\mcitedefaultseppunct}\relax
	\EndOfBibitem
	\bibitem[Huang \emph{et~al.}(2017)Huang, Clark, Navarro-Moratalla, Klein,
	Cheng, Seyler, Zhong, Schmidgall, McGuire,
	Cobden,\emph{et~al.}]{huang2017layer}
	B.~Huang, G.~Clark, E.~Navarro-Moratalla, D.~R. Klein, R.~Cheng, K.~L. Seyler,
	D.~Zhong, E.~Schmidgall, M.~A. McGuire, D.~H. Cobden \emph{et~al.},
	\emph{Nature}, 2017, \textbf{546}, 270\relax
	\mciteBstWouldAddEndPuncttrue
	\mciteSetBstMidEndSepPunct{\mcitedefaultmidpunct}
	{\mcitedefaultendpunct}{\mcitedefaultseppunct}\relax
	\EndOfBibitem
	\bibitem[Gong \emph{et~al.}(2017)Gong, Li, Li, Ji, Stern, Xia, Cao, Bao, Wang,
	Wang,\emph{et~al.}]{gong2017discovery}
	C.~Gong, L.~Li, Z.~Li, H.~Ji, A.~Stern, Y.~Xia, T.~Cao, W.~Bao, C.~Wang,
	Y.~Wang \emph{et~al.}, \emph{Nature}, 2017, \textbf{546}, 265\relax
	\mciteBstWouldAddEndPuncttrue
	\mciteSetBstMidEndSepPunct{\mcitedefaultmidpunct}
	{\mcitedefaultendpunct}{\mcitedefaultseppunct}\relax
	\EndOfBibitem
	\bibitem[Gr{\"o}nke \emph{et~al.}(2019)Gr{\"o}nke, Buschbeck, Schmidt, Valldor,
	Oswald, Hao, Lubk, Wolf, Steiner,
	B{\"u}chner,\emph{et~al.}]{gronke2019chromium}
	M.~Gr{\"o}nke, B.~Buschbeck, P.~Schmidt, M.~Valldor, S.~Oswald, Q.~Hao,
	A.~Lubk, D.~Wolf, U.~Steiner, B.~B{\"u}chner \emph{et~al.}, \emph{Advanced
		materials interfaces}, 2019, \textbf{6}, 1901410\relax
	\mciteBstWouldAddEndPuncttrue
	\mciteSetBstMidEndSepPunct{\mcitedefaultmidpunct}
	{\mcitedefaultendpunct}{\mcitedefaultseppunct}\relax
	\EndOfBibitem
	\bibitem[Wang \emph{et~al.}(2018)Wang, Zhang, Ding, Dong, Li, Chen, Li, Huang,
	Wang, Zhao,\emph{et~al.}]{wang2018electric}
	Z.~Wang, T.~Zhang, M.~Ding, B.~Dong, Y.~Li, M.~Chen, X.~Li, J.~Huang, H.~Wang,
	X.~Zhao \emph{et~al.}, \emph{Nature nanotechnology}, 2018, \textbf{13},
	554\relax
	\mciteBstWouldAddEndPuncttrue
	\mciteSetBstMidEndSepPunct{\mcitedefaultmidpunct}
	{\mcitedefaultendpunct}{\mcitedefaultseppunct}\relax
	\EndOfBibitem
	\bibitem[Sun \emph{et~al.}(2020)Sun, Li, Wang, Sui, Zhang, Wang, Liu, Li, Feng,
	Zhong,\emph{et~al.}]{sun2020room}
	X.~Sun, W.~Li, X.~Wang, Q.~Sui, T.~Zhang, Z.~Wang, L.~Liu, D.~Li, S.~Feng,
	S.~Zhong \emph{et~al.}, \emph{Nano Research}, 2020, \textbf{13}, 3358\relax
	\mciteBstWouldAddEndPuncttrue
	\mciteSetBstMidEndSepPunct{\mcitedefaultmidpunct}
	{\mcitedefaultendpunct}{\mcitedefaultseppunct}\relax
	\EndOfBibitem
	\bibitem[Gong \emph{et~al.}(2019)Gong, Guo, Li, Zhu, Liao, Liu, Zhang, Gu,
	Tang, Feng,\emph{et~al.}]{gong2019experimental}
	Y.~Gong, J.~Guo, J.~Li, K.~Zhu, M.~Liao, X.~Liu, Q.~Zhang, L.~Gu, L.~Tang,
	X.~Feng \emph{et~al.}, \emph{Chinese Physics Letters}, 2019, \textbf{36},
	076801\relax
	\mciteBstWouldAddEndPuncttrue
	\mciteSetBstMidEndSepPunct{\mcitedefaultmidpunct}
	{\mcitedefaultendpunct}{\mcitedefaultseppunct}\relax
	\EndOfBibitem
	\bibitem[Huang \emph{et~al.}(2018)Huang, Clark, Klein, MacNeill,
	Navarro-Moratalla, Seyler, Wilson, McGuire, Cobden, Xiao, Yao,
	Jarillo-Herrero, and Xu]{cri3}
	B.~Huang, G.~Clark, D.~R. Klein, D.~MacNeill, E.~Navarro-Moratalla, K.~L.
	Seyler, N.~Wilson, M.~A. McGuire, D.~H. Cobden, D.~Xiao, W.~Yao,
	P.~Jarillo-Herrero and X.~Xu, \emph{Nature Nanotechnology}, 2018,
	\textbf{13}, 544\relax
	\mciteBstWouldAddEndPuncttrue
	\mciteSetBstMidEndSepPunct{\mcitedefaultmidpunct}
	{\mcitedefaultendpunct}{\mcitedefaultseppunct}\relax
	\EndOfBibitem
	\bibitem[Jiang \emph{et~al.}(2018)Jiang, Li, Wang, Mark, and Shan]{cri31}
	S.~Jiang, L.~Li, Z.~Wang, K.~F. Mark and J.~Shan, \emph{Nature Nanotechnology},
	2018, \textbf{13}, 549\relax
	\mciteBstWouldAddEndPuncttrue
	\mciteSetBstMidEndSepPunct{\mcitedefaultmidpunct}
	{\mcitedefaultendpunct}{\mcitedefaultseppunct}\relax
	\EndOfBibitem
	\bibitem[Jiang \emph{et~al.}(2018)Jiang, Shan, and Mark]{cri32}
	S.~Jiang, J.~Shan and K.~F. Mark, \emph{Nature Materials}, 2018, \textbf{17},
	406\relax
	\mciteBstWouldAddEndPuncttrue
	\mciteSetBstMidEndSepPunct{\mcitedefaultmidpunct}
	{\mcitedefaultendpunct}{\mcitedefaultseppunct}\relax
	\EndOfBibitem
	\bibitem[Wang \emph{et~al.}(2018)Wang, Zhang, Ding, Dong, Li, Chen, Li, Huang,
	Wang, Zhao, Li, Li, Jia, Sun, Guo, Ye, Sun, Chen, Yang, Zhang, Ono, Han, and
	Zhang]{crgete}
	J.~Wang, T.~Zhang, M.~Ding, B.~Dong, Y.~Li, M.~Chen, X.~Li, J.~Huang, H.~Wang,
	X.~Zhao, Y.~Li, D.~Li, C.~Jia, L.~Sun, H.~Guo, Y.~Ye, D.~Sun, Y.~Chen,
	T.~Yang, J.~Zhang, S.~Ono, Z.~Han and Z.~Zhang, \emph{Nature Nanotechnology},
	2018, \textbf{13}, 554\relax
	\mciteBstWouldAddEndPuncttrue
	\mciteSetBstMidEndSepPunct{\mcitedefaultmidpunct}
	{\mcitedefaultendpunct}{\mcitedefaultseppunct}\relax
	\EndOfBibitem
	\bibitem[Deng \emph{et~al.}(2018)Deng, Yu, Song, Zhang, Wang, Sun, Yi, Wu, Wu,
	Zhu, Wang, Chen, and Zhang]{fegete}
	Y.~Deng, Y.~Yu, Y.~Song, J.~Zhang, N.~Z. Wang, Z.~Sun, Y.~Yi, Y.~Z. Wu, S.~Wu,
	J.~Zhu, J.~Wang, X.~H. Chen and Y.~Zhang, \emph{Nature}, 2018, \textbf{563},
	94\relax
	\mciteBstWouldAddEndPuncttrue
	\mciteSetBstMidEndSepPunct{\mcitedefaultmidpunct}
	{\mcitedefaultendpunct}{\mcitedefaultseppunct}\relax
	\EndOfBibitem
	\bibitem[Fei \emph{et~al.}(2018)Fei, Huang, Malinowski, Wang, Song, Sanchez,
	Yao, Xiao, Zhu, May, Wu, Cobden, Chu, and Xu]{fegete1}
	Z.~Fei, B.~Huang, P.~Malinowski, W.~Wang, T.~Song, J.~Sanchez, W.~Yao, D.~Xiao,
	X.~Zhu, A.~F. May, W.~Wu, D.~H. Cobden, J.~H. Chu and X.~Xu, \emph{Nature
		Materials}, 2018, \textbf{17}, 778\relax
	\mciteBstWouldAddEndPuncttrue
	\mciteSetBstMidEndSepPunct{\mcitedefaultmidpunct}
	{\mcitedefaultendpunct}{\mcitedefaultseppunct}\relax
	\EndOfBibitem
	\bibitem[Dong \emph{et~al.}(2019)Dong, Xiang, and
	Dagotto]{dong2019magnetoelectricity}
	S.~Dong, H.~Xiang and E.~Dagotto, \emph{National Science Review}, 2019,
	\textbf{6}, 629\relax
	\mciteBstWouldAddEndPuncttrue
	\mciteSetBstMidEndSepPunct{\mcitedefaultmidpunct}
	{\mcitedefaultendpunct}{\mcitedefaultseppunct}\relax
	\EndOfBibitem
	\bibitem[Gong \emph{et~al.}(2019)Gong, Kim, Wang, Lee, and
	Zhang]{gong2019multiferroicity}
	C.~Gong, E.~M. Kim, Y.~Wang, G.~Lee and X.~Zhang, \emph{Nature Communications},
	2019, \textbf{10}, 2657\relax
	\mciteBstWouldAddEndPuncttrue
	\mciteSetBstMidEndSepPunct{\mcitedefaultmidpunct}
	{\mcitedefaultendpunct}{\mcitedefaultseppunct}\relax
	\EndOfBibitem
	\bibitem[Sun \emph{et~al.}(2019)Sun, Wang, Chen, Cheng, and
	Wang]{sun2019valence}
	W.~Sun, W.~Wang, D.~Chen, Z.~Cheng and Y.~Wang, \emph{Nanoscale}, 2019,
	\textbf{11}, 9931\relax
	\mciteBstWouldAddEndPuncttrue
	\mciteSetBstMidEndSepPunct{\mcitedefaultmidpunct}
	{\mcitedefaultendpunct}{\mcitedefaultseppunct}\relax
	\EndOfBibitem
	\bibitem[Spaldin and Fiebig(2005)]{sciencearticle}
	N.~A. Spaldin and M.~Fiebig, \emph{Science}, 2005, \textbf{309}, 391\relax
	\mciteBstWouldAddEndPuncttrue
	\mciteSetBstMidEndSepPunct{\mcitedefaultmidpunct}
	{\mcitedefaultendpunct}{\mcitedefaultseppunct}\relax
	\EndOfBibitem
	\bibitem[Wan \emph{et~al.}(2019)Wan, Li, Li, Mao, Wang, Chen, Dong, Nie, Xiang,
	Liu,\emph{et~al.}]{wan2019nonvolatile}
	S.~Wan, Y.~Li, W.~Li, X.~Mao, C.~Wang, C.~Chen, J.~Dong, A.~Nie, J.~Xiang,
	Z.~Liu \emph{et~al.}, \emph{Advanced Functional Materials}, 2019,
	\textbf{29}, 1808606\relax
	\mciteBstWouldAddEndPuncttrue
	\mciteSetBstMidEndSepPunct{\mcitedefaultmidpunct}
	{\mcitedefaultendpunct}{\mcitedefaultseppunct}\relax
	\EndOfBibitem
	\bibitem[Wang \emph{et~al.}(2020)Wang, Wang, Zhang, Li, Ma, Leng, Chen,
	Abdelwahab, and Loh]{wang2020exploring}
	L.~Wang, X.~Wang, Y.~Zhang, R.~Li, T.~Ma, K.~Leng, Z.~Chen, I.~Abdelwahab and
	K.~P. Loh, \emph{Advanced Functional Materials}, 2020, \textbf{30},
	2004609\relax
	\mciteBstWouldAddEndPuncttrue
	\mciteSetBstMidEndSepPunct{\mcitedefaultmidpunct}
	{\mcitedefaultendpunct}{\mcitedefaultseppunct}\relax
	\EndOfBibitem
	\bibitem[Yu \emph{et~al.}(2018)Yu, Gao, Wang, Du, Lin, Guo, Zou, Jin, Li, and
	Chen]{yu2018unraveling}
	H.~Yu, D.~Gao, X.~Wang, X.~Du, X.~Lin, W.~Guo, R.~Zou, C.~Jin, K.~Li and
	Y.~Chen, \emph{NPG Asia Materials}, 2018, \textbf{10}, 882\relax
	\mciteBstWouldAddEndPuncttrue
	\mciteSetBstMidEndSepPunct{\mcitedefaultmidpunct}
	{\mcitedefaultendpunct}{\mcitedefaultseppunct}\relax
	\EndOfBibitem
	\bibitem[Chang \emph{et~al.}(2016)Chang, Liu, Lin, Wang, Zhao, Zhang, Jin,
	Zhong, Hu, Duan,\emph{et~al.}]{chang2016discovery}
	K.~Chang, J.~Liu, H.~Lin, N.~Wang, K.~Zhao, A.~Zhang, F.~Jin, Y.~Zhong, X.~Hu,
	W.~Duan \emph{et~al.}, \emph{Science}, 2016, \textbf{353}, 274\relax
	\mciteBstWouldAddEndPuncttrue
	\mciteSetBstMidEndSepPunct{\mcitedefaultmidpunct}
	{\mcitedefaultendpunct}{\mcitedefaultseppunct}\relax
	\EndOfBibitem
	\bibitem[Liu \emph{et~al.}(2016)Liu, You, Seyler, Li, Yu, Lin, Wang, Zhou,
	Wang, He,\emph{et~al.}]{liu2016room}
	F.~Liu, L.~You, K.~L. Seyler, X.~Li, P.~Yu, J.~Lin, X.~Wang, J.~Zhou, H.~Wang,
	H.~He \emph{et~al.}, \emph{Nature Communications}, 2016, \textbf{7}, 1\relax
	\mciteBstWouldAddEndPuncttrue
	\mciteSetBstMidEndSepPunct{\mcitedefaultmidpunct}
	{\mcitedefaultendpunct}{\mcitedefaultseppunct}\relax
	\EndOfBibitem
	\bibitem[Noor-A-Alam and Nolan(2023)]{noor2023engineering}
	M.~Noor-A-Alam and M.~Nolan, \emph{ACS Applied Materials \& Interfaces}, 2023,
	\textbf{15}, 42737\relax
	\mciteBstWouldAddEndPuncttrue
	\mciteSetBstMidEndSepPunct{\mcitedefaultmidpunct}
	{\mcitedefaultendpunct}{\mcitedefaultseppunct}\relax
	\EndOfBibitem
	\bibitem[Kohn and Sham(1965)]{dft}
	W.~Kohn and L.~J. Sham, \emph{Physical Review}, 1965, \textbf{140}, A1133\relax
	\mciteBstWouldAddEndPuncttrue
	\mciteSetBstMidEndSepPunct{\mcitedefaultmidpunct}
	{\mcitedefaultendpunct}{\mcitedefaultseppunct}\relax
	\EndOfBibitem
	\bibitem[Zhao \emph{et~al.}(2019)Zhao, Zhang, Yuan, and
	Chen]{zhao2019nonvolatile}
	Y.~Zhao, J.-J. Zhang, S.~Yuan and Z.~Chen, \emph{Advanced Functional
		Materials}, 2019, \textbf{29}, 1901420\relax
	\mciteBstWouldAddEndPuncttrue
	\mciteSetBstMidEndSepPunct{\mcitedefaultmidpunct}
	{\mcitedefaultendpunct}{\mcitedefaultseppunct}\relax
	\EndOfBibitem
	\bibitem[Liu \emph{et~al.}(2023)Liu, Chen, Zhou, Xu, and
	Xiao]{liu2023nonvolatile}
	G.~Liu, T.~Chen, G.~Zhou, Z.~Xu and X.~Xiao, \emph{ACS sensors}, 2023,
	\textbf{8}, 1440--1449\relax
	\mciteBstWouldAddEndPuncttrue
	\mciteSetBstMidEndSepPunct{\mcitedefaultmidpunct}
	{\mcitedefaultendpunct}{\mcitedefaultseppunct}\relax
	\EndOfBibitem
	\bibitem[Wu \emph{et~al.}(2024)Wu, Sun, Gong, Li, and Wang]{wu2024nonvolatile}
	C.~Wu, S.~Sun, W.~Gong, J.~Li and X.~Wang, \emph{Physical Chemistry Chemical
		Physics}, 2024, \textbf{26}, 5323\relax
	\mciteBstWouldAddEndPuncttrue
	\mciteSetBstMidEndSepPunct{\mcitedefaultmidpunct}
	{\mcitedefaultendpunct}{\mcitedefaultseppunct}\relax
	\EndOfBibitem
	\bibitem[Cao \emph{et~al.}(2022)Cao, Deng, Zhou, Liang, Nguyen, Ang, and
	Ang]{cao2022multiferroic}
	L.~Cao, X.~Deng, G.~Zhou, S.-J. Liang, C.~V. Nguyen, L.~Ang and Y.~S. Ang,
	\emph{Physical Review B}, 2022, \textbf{105}, 165302\relax
	\mciteBstWouldAddEndPuncttrue
	\mciteSetBstMidEndSepPunct{\mcitedefaultmidpunct}
	{\mcitedefaultendpunct}{\mcitedefaultseppunct}\relax
	\EndOfBibitem
	\bibitem[Song \emph{et~al.}(2024)Song, Ye, Su, Wei, Chen, Liu, Zheng, and
	Hao]{vse2}
	L.~Song, R.~Ye, C.~Su, C.~Wei, D.~Chen, X.~Liu, X.~Zheng and H.~Hao,
	\emph{Physical Review B}, 2024, \textbf{109}, 094105\relax
	\mciteBstWouldAddEndPuncttrue
	\mciteSetBstMidEndSepPunct{\mcitedefaultmidpunct}
	{\mcitedefaultendpunct}{\mcitedefaultseppunct}\relax
	\EndOfBibitem
	\bibitem[Liu and Ke(2024)]{maz}
	G.~Liu and S.~H. Ke, \emph{Physical Review Applied}, 2024, \textbf{21},
	044033\relax
	\mciteBstWouldAddEndPuncttrue
	\mciteSetBstMidEndSepPunct{\mcitedefaultmidpunct}
	{\mcitedefaultendpunct}{\mcitedefaultseppunct}\relax
	\EndOfBibitem
	\bibitem[Chandrasekaran \emph{et~al.}(2017)Chandrasekaran, Mishra, and
	Singh]{sc2co2abhishek}
	A.~Chandrasekaran, A.~Mishra and A.~K. Singh, \emph{Nano Letters}, 2017,
	\textbf{17}, 3290\relax
	\mciteBstWouldAddEndPuncttrue
	\mciteSetBstMidEndSepPunct{\mcitedefaultmidpunct}
	{\mcitedefaultendpunct}{\mcitedefaultseppunct}\relax
	\EndOfBibitem
	\bibitem[You \emph{et~al.}(2022)You, Pan, Shang, Xu, Liu, Li, Liu, Kang, Xu,
	Li,\emph{et~al.}]{you2022salt}
	J.~You, J.~Pan, S.-L. Shang, X.~Xu, Z.~Liu, J.~Li, H.~Liu, T.~Kang, M.~Xu,
	S.~Li \emph{et~al.}, \emph{Nano Letters}, 2022, \textbf{22}, 10167\relax
	\mciteBstWouldAddEndPuncttrue
	\mciteSetBstMidEndSepPunct{\mcitedefaultmidpunct}
	{\mcitedefaultendpunct}{\mcitedefaultseppunct}\relax
	\EndOfBibitem
	\bibitem[Kresse and Furthm{\"u}ller(1996)]{kresse1996efficient}
	G.~Kresse and J.~Furthm{\"u}ller, \emph{Physical Review B}, 1996, \textbf{54},
	11169\relax
	\mciteBstWouldAddEndPuncttrue
	\mciteSetBstMidEndSepPunct{\mcitedefaultmidpunct}
	{\mcitedefaultendpunct}{\mcitedefaultseppunct}\relax
	\EndOfBibitem
	\bibitem[Kresse and Joubert(1999)]{kresse1999ultrasoft}
	G.~Kresse and D.~Joubert, \emph{Physical Review B}, 1999, \textbf{59},
	1758\relax
	\mciteBstWouldAddEndPuncttrue
	\mciteSetBstMidEndSepPunct{\mcitedefaultmidpunct}
	{\mcitedefaultendpunct}{\mcitedefaultseppunct}\relax
	\EndOfBibitem
	\bibitem[Perdew \emph{et~al.}(1996)Perdew, Burke, and
	Ernzerhof]{perdew1996generalized}
	J.~P. Perdew, K.~Burke and M.~Ernzerhof, \emph{Physical review letters}, 1996,
	\textbf{77}, 3865\relax
	\mciteBstWouldAddEndPuncttrue
	\mciteSetBstMidEndSepPunct{\mcitedefaultmidpunct}
	{\mcitedefaultendpunct}{\mcitedefaultseppunct}\relax
	\EndOfBibitem
	\bibitem[Grimme \emph{et~al.}(2010)Grimme, Antony, Ehrlich, and
	Krieg]{grimme2010consistent}
	S.~Grimme, J.~Antony, S.~Ehrlich and H.~Krieg, \emph{The Journal of chemical
		physics}, 2010, \textbf{132}, 154104\relax
	\mciteBstWouldAddEndPuncttrue
	\mciteSetBstMidEndSepPunct{\mcitedefaultmidpunct}
	{\mcitedefaultendpunct}{\mcitedefaultseppunct}\relax
	\EndOfBibitem
	\bibitem[Monkhorst and Pack(1976)]{mp}
	H.~J. Monkhorst and J.~D. Pack, \emph{Phyical Review B}, 1976, \textbf{13},
	5188\relax
	\mciteBstWouldAddEndPuncttrue
	\mciteSetBstMidEndSepPunct{\mcitedefaultmidpunct}
	{\mcitedefaultendpunct}{\mcitedefaultseppunct}\relax
	\EndOfBibitem
	\bibitem[Wang \emph{et~al.}(2021)Wang, Li, Li, Wu, Che, Chen, and
	Cui]{wang2021ferromagnetism}
	X.~Wang, D.~Li, Z.~Li, C.~Wu, C.-M. Che, G.~Chen and X.~Cui, \emph{ACS Nano},
	2021, \textbf{15}, 16236\relax
	\mciteBstWouldAddEndPuncttrue
	\mciteSetBstMidEndSepPunct{\mcitedefaultmidpunct}
	{\mcitedefaultendpunct}{\mcitedefaultseppunct}\relax
	\EndOfBibitem
	\bibitem[Fuh \emph{et~al.}(2016)Fuh, Chang, Wang, Evans, Chantrell, and
	Jeng]{fuh2016newtype}
	H.-R. Fuh, C.-R. Chang, Y.-K. Wang, R.~F. Evans, R.~W. Chantrell and H.-T.
	Jeng, \emph{Scientific Reports}, 2016, \textbf{6}, 32625\relax
	\mciteBstWouldAddEndPuncttrue
	\mciteSetBstMidEndSepPunct{\mcitedefaultmidpunct}
	{\mcitedefaultendpunct}{\mcitedefaultseppunct}\relax
	\EndOfBibitem
	\bibitem[Wang \emph{et~al.}(2023)Wang, Qiu, Wu, Huang, An, He, and
	Du]{wang2023first}
	Z.~Wang, N.~Qiu, E.~Wu, Q.~Huang, P.~An, H.~He and S.~Du, \emph{Journal of
		Materials Research and Technology}, 2023, \textbf{24}, 173\relax
	\mciteBstWouldAddEndPuncttrue
	\mciteSetBstMidEndSepPunct{\mcitedefaultmidpunct}
	{\mcitedefaultendpunct}{\mcitedefaultseppunct}\relax
	\EndOfBibitem
	\bibitem[Li \emph{et~al.}(2012)Li, Wu, Li, Yang, and Hou]{li2012bipolar}
	X.~Li, X.~Wu, Z.~Li, J.~Yang and J.~Hou, \emph{Nanoscale}, 2012, \textbf{4},
	5680\relax
	\mciteBstWouldAddEndPuncttrue
	\mciteSetBstMidEndSepPunct{\mcitedefaultmidpunct}
	{\mcitedefaultendpunct}{\mcitedefaultseppunct}\relax
	\EndOfBibitem
	\bibitem[Zhang \emph{et~al.}(2020)Zhang, Tang, Zhang, and Du]{mxenefe}
	L.~Zhang, C.~Tang, C.~Zhang and A.~Du, \emph{Nanoscale}, 2020, \textbf{12},
	21291\relax
	\mciteBstWouldAddEndPuncttrue
	\mciteSetBstMidEndSepPunct{\mcitedefaultmidpunct}
	{\mcitedefaultendpunct}{\mcitedefaultseppunct}\relax
	\EndOfBibitem
\end{mcitethebibliography}
\providecommand*{\mcitethebibliography}{\thebibliography}
\csname @ifundefined\endcsname{endmcitethebibliography}
{\let\endmcitethebibliography\endthebibliography}{}

\end{document}